\documentclass[12pt,preprint]{revtex4-1}
\usepackage{graphicx} % Required for inserting images
\usepackage{epsfig}  
\usepackage{amssymb,amsmath}
\usepackage{tabularx}
\usepackage[table,xcdraw]{xcolor}
\usepackage{natbib}
\usepackage[greek,english]{babel}
\usepackage{subcaption}
\usepackage{float}
\usepackage[colorlinks=true,linkcolor=blue,urlcolor=blue]
{hyperref}

\usepackage{caption}
\usepackage{svg}
\usepackage[normalem]{ulem}
\usepackage{soul}
\sethlcolor{yellow}

\begin{document}

\title{Can rising consumption deepen inequality?}

\author{Jhordan Silveira de Borba}
\email{sbjhordan@gmail.com}
\affiliation{Instituto de Física, Universidade Federal do Rio Grande do Sul, Av. Bento Gonçalves 9500, Porto Alegre, 90560-001, RS, Brazil}

\author{Celia Anteneodo}
\email{celia.fis@puc-rio.br}
\affiliation{Departamento de Física, PUC-Rio, Rua Marquês de São Vicente 225, Rio de Janeiro, 22451-900, RJ, Brazil}

\author{Sebastian Gonçalves}
\email{sebastiangoncalves@gmail.com}
\affiliation{Instituto de Fisica, Universidade Federal do Rio Grande do Sul, Av. Bento Gonçalves 9500, Porto Alegre, 90560-001, RS, Brazil}

\begin{abstract}
The impact of rising consumption on wealth inequality remains an open question. Here we revisit and extend the Social Architecture of Capitalism agent-based model proposed by Ian Wright, which reproduces stylized facts of wealth and income distributions. In a previous study, we demonstrated that the macroscopic behavior of the model is predominantly governed by a single dimensionless parameter, the ratio between average wealth per capita and mean salary, denoted by $R$. The shape of the wealth distribution, the emergence of a two-class structure, and the level of inequality ---summarized by the Gini index--- were found to depend mainly on $R$, with inequality increasing as $R$ increases.
In the present work, we examine the robustness of this result by relaxing some simplifying assumptions of the model. We first allow transactions such as purchases, salary payments, and revenue collections to occur with different frequencies, reflecting the heterogeneous temporal dynamics of real economies. We then impose limits on the maximum fractions of wealth that agents can spend or collect at each step, constraining the amplitude of individual transactions. We find that the dependence of the inequality on $R$ remains qualitatively robust, although the detailed distribution patterns are affected by relative frequencies and transaction limits.
Finally, we analyze a further variant of the model with adaptive wages emerging endogenously from the dynamics, showing that self-organized labor-market feedback can either stabilize or amplify inequality depending on macroeconomic conditions.

\end{abstract}

\keywords{Econophysics, Wealth Distribution, Inequality, Risk, Agent-based simulations}

%\date{\today}

\maketitle

\section{Introduction} 
\label{sec:introduction}

  Economic systems can be viewed as complex systems whose macroscopic properties emerge from interactions among many heterogeneous agents~\cite{JUSUP20221,yakovenko2009,Farmer2009}. This perspective, developed in econophysics, has provided important insights into empirical regularities, particularly in the distribution of income and wealth~\cite{shaikh}. 

In fact, in econophysics, a central problem is to identify the mechanisms responsible for the emergence of highly unequal income and wealth distributions, typically characterized by broad, often heavy-tailed forms observed across different economies and time periods. 
 While only a limited number of modeling approaches can reproduce these distributions, often capturing only specific features, identifying the underlying mechanisms responsible for their robustness and universality remains a major challenge. 

  Motivated by this approach, a variety of models have been proposed to describe the dynamics of wealth and income, including kinetic exchange models and agent-based frameworks. 
Within this class of models, the Social Architecture of Capitalism (SA) model~\cite{WRIGHT2005589}, proposed by Ian Wright, offers a simple yet realistic representation of economic interactions between agents, reproducing key empirical features of wealth distributions.  

 In a recent work~\cite{Borba2025}, we revisited the SA model and systematically explored a wide range of its parameters. A two-class structure of wealth distribution emerges, producing an upper tail following a Pareto power-law~\cite{dragulescu2001,yakovenko2023}, and a lower segment characterized by an exponential or lognormal form. We identified a key parameter that is responsible for opening a gap between the two groups.

This key parameter, denoted by $R$, is defined as the ratio between the average wealth per capita and the mean salary. 
We showed that the resulting wealth distribution, its characteristic patterns, the eventual separation of wealth, and the extinction of the middle class for large values of $R$, depend primarily on this parameter, with only a weak dependence on system size. 
In particular, inequality as summarized by the Gini index~\cite{PLATAPEREZ201579}, was found to increase  monotonically with $R$.  
This prediction of the model is consistent with results obtained from  empirical data in which $\overline{p}$ and  $\overline{w}$ are estimated using  average employee compensation and  net personal wealth, respectively, based, among other sources, on data from  the World Inequality Database (WID) \cite{WID}.
 
However, the model relies on strong simplifying assumptions that may oversimplify the complex dynamics of real economies. For instance, in the model all transactions (salary payments, purchases, and revenue collections) occur with the same frequency, i.e., at every time step. In real economies, however, purchases typically occur more frequently than the reception of salaries or the collection of revenues. We therefore aim to investigate how relaxing this rule, by introducing different frequencies for these processes relative to salary periodicity, may affect the results.

A similar question arises regarding the range of possible expenses (or revenue collections), which in the original formulation can take any value between zero and the agent’s total wealth (or total market value). In other words, agents are allowed to spend their entire wealth (or collect the entire market value) in a single step. Here, we explore how limiting these maxima—both expenditures and market value collections—affects the model’s dynamics and the resulting inequality patterns.

Furthermore, we sought to investigate how the system behaves in alternative formulations in which the average wage is no longer fixed exogenously but instead emerges endogenously from the internal model dynamics.

The model is defined in Section~\ref{sec:model}, with emphasis on the introduced variations. Section~\ref{sec:results} presents the results of the agent-based simulations, examining how consumption rules, market revenues, and employment regulations affect inequality. Finally, Section~\ref{sec:final} provides the conclusions and final remarks.

\section{Model}
\label{sec:model}

The agent-based model implemented in this paper introduces variations in the original SA model proposed by Ian Wright~\cite{WRIGHT2005589}. 

This model is based on the assumption that `social relations of production' \cite{marx2019capital} define the roles that individuals play in economic activity. In particular, the employer–employee relationship mediated by wage labor is taken as an intrinsic feature of capitalism. In this context, the structure of inequality emerges from the organization of these relationships, insofar as a small class of capitalists receives its income in the form of profits, while a large class of workers has wages as its sole source of income.

This approach allows one to abstract from specific and often transitory characteristics of concrete capitalist economies, such as natural or technical constraints or particular types of commodities, and to focus on interactions between agents mediated by monetary exchange. The model thus reduces the economy to a minimal structure in which individuals, differentiated by their position in the relations of production, interact through money flows,  and this dynamics is responsible for generating the patterns of inequality observed.

The society consists of $N$ agents (individuals or other economic entities, identified by the label $i=1,\dots,N$). Each agent $i$ has a quantity $w_i(t)$ representing wealth, which varies over time. 
Although in the original proposal money takes integer values 
as multiples of a monetary unit, we have verified that no significant differences are observed when considering  real values, which is the choice in the present work.

The size of the population $N$ and the total wealth of the system are conserved, so the wealth per capita $\overline{w}=W/N$ is constant.
At any given time $t$, agents belong to one of three classes: employees (working class), employers (capitalist class), or unemployed. The class to which each agent belongs is dynamically updated. 
Then each  agent is characterized by an index $e_i$ that identifies their employer: $e_i=j$ indicates that agent $j$ is the employer of agent $i$. There are two special cases: $e_i=0$ indicates that the agent is unemployed and $e_i=-1$ indicates that they are an employer.  
Thus, at any time $t$, the state of the entire economy is defined by the set of pairs $S(t) = \left\{ \left( w_i(t), e_i(t) \right) :1\leq i\leq N \right\}$. It is important to note that a firm consists of a set of employees and a single  employer who is the sole owner of the firm, and there is no self-employment.

In addition to agents, the model includes a market entity that represents a monetary pool with value $V$ and evolves through transactions. Consumer expenditures increase $V$, while firm revenues decrease it, so firms compete for shares of this pool rather than being paid directly by individual buyers.

In its original formulation, the model is specified by four parameters: the number of agents $N,$ the bounds $(p_a,p_b)$ of the wage distribution, and the total money supply $W,$ which fixes the initial average wealth per capita $\overline{w} = W/N$. In practice, however, we have shown that the macroscopic behavior of the system does not depend on the individual values of $p_a$ and $p_b$, but only on their mean $\overline{p}=(p_a+p_b)/2$~\cite{Borba2025}. We therefore fix a constant unit wage, $\overline{p}=p$.
Since $\overline{p}$ sets the monetary unit, the only remaining intensive control parameter is the ratio $R=\overline{w}/\overline{p}$, which measures the wealth per capita in units of the average wage. The model thus depends effectively on $R$, with a weak residual dependence on the system size $N$. In practice, no significant finite-size effects are observed for $N \gtrsim 20$.

In what follows, we summarize the original rules, emphasizing the modifications. In the initial state, all agents have the same wealth ($w_i(0)=\overline{w}$) and the same state ($e_i(0)=0$), and the following set of five rules is applied, in the given order, after randomly selecting an agent $i$.

\begin{itemize}

\item  {\bf Hiring} -- If $i$ is unemployed ($e_i=0$), a potential employer $j$ is selected from the set $H$  of all agents who are not employees ($e_j \leq 0$), with  probability  $P (j)= w_{j}/\sum_{n\in H}w_{n}$.   If $ w_j \ge p$, agent $i$ is hired by $j$ (thus, $e_i=j$).    If $j$ was previously unemployed, $j$ becomes an employer ($e_j=-1$).

\item   {\bf  Expenditure} (on goods and services) -- A random agent $j$ is selected, and an amount $w$, representing the agent's expenditure, is subtracted from their wealth ($w_j \to w_j-w$) and added to the market value ($V \to V+w$).
We modify this rule, introducing two new parameters,  $\Omega_E$ associated with the frequency of visits to the market, and $\Phi_E$ representing the fraction of wealth spent in the market.
If $\Omega_E\leq1$, it denotes the probability that this rule will be applied.  
Otherwise, $\Omega_E$ indicates the number of times the rule is applied. 
For each application, a random agent $j$ is selected and a random expenditure value $w$ is drawn from the range $\left[0,\Phi_E w_j\right]$. 
 
\item {\bf Market revenue} (from sales of goods and services) -- 
A random agent $j$ is selected, and an amount $w$ is  transferred from the market value ($V \to V - w$) to the agent $j$.  If $j$ is an employer, the amount is credited to $j$ ($w_j \to w_j + w$), if $j$ is an employee, it is credited to the employer $e_j$  ($w_{e_j} \to w_{e_j} + w$).  In all cases,  $w$ represents revenue for the owner of the company.
We modify this rule, introducing two new parameters, namely,  $\Omega_M$, associated with the frequency of visits to the market, and $\Phi_M$, representing the fraction of wealth withdrawn from the market value.
If $\Omega_M\leq1$, it denotes the probability that this rule will be applied.
Otherwise, $\Omega_M$ specifies the number of times the rule is applied.  For each application, a new random agent $j$ is selected; if $j$  is not unemployed ($e_i\neq 0$), a random value is drawn from the interval $w \in [0, \Phi_M V]$.

\item {\bf Firing --} If agent $i$ acts as an employer ($e_i=-1$), its financial constraint determines how many employees must be dismissed. Let $n_i$ be the number of agents employed by $i$ (i.e., those with $e_j=i$). Since each worker is paid a wage $p$, the employer can afford at most $w_i/p$ employees. If $n_i > w_i/p$, the excess number of workers
\begin{equation*}
n = \max\!\left(n_i - \frac{w_i}{p},\,0\right)
\end{equation*}
is dismissed. The $n$ agents to be fired are selected uniformly at random from $i$’s list of employees. If all employees are dismissed, agent $i$ becomes unemployed ($e_i=0$).

\item {\bf Wage payment --} 
For each employee $j$ of an employer $i$ (i.e., $e_j=i$), the wage $p$ is paid by $i$, resulting in the updates $w_i \to w_i - p$ and $w_j \to w_j + p$.
\end{itemize}

%%%%
The update is asynchronous, meaning that the state of the system is updated after each individual rule is applied. A sequence of $N$ rule applications (one Monte Carlo step) corresponds, arbitrarily, to one month, giving each agent, on average, the opportunity to be selected once per month. We verified that no significant differences arise when changing the order of the rules; therefore, they are applied in random order.

This set of rules, including the modifications introduced in the {\it Expenditure} and {\it Market revenue} rules that govern market exchanges, is used to obtain the results presented in Sections~\ref{sec:expenditure} and~\ref{sec:revenue}. 
The code is publicly available\cite{code}.

Below, we consider an alternative variant of the model obtained by modifying the rules related to employment. This version is used exclusively for the results presented in Section~\ref{sec:hiring}.

\subsection*{Modifications in employment rules}

\label{modifications}

%%%
We propose a modification of the {\it Hiring} rule in which wages are no longer fixed but instead emerge from the system dynamics. Furthermore, agents are allowed to seek alternative employment offering higher wages even while already employed, a realistic feature absent in the original model.

By removing the fixed-wage constraint, the model captures a more dynamic labor market with varying wage levels. The rationale for this modification is as follows: unemployed workers seek employment, and the longer they remain unemployed, the more they  lower their wage expectations, reflecting a weakened bargaining position. Conversely, employed agents search for alternative opportunities offering higher wages than their current ones.

Under this rule, in contrast to the fixed-wage model, both wages and expectations adjust endogenously to labor market conditions through a feedback mechanism. 
This coupling between labor market conditions and wage formation drives the evolution of income and wealth distributions over time.

To enable this feature, we introduce two agent-specific variables and one system parameter. The variables are $\mu_i$, which records the last wage received by agent $i$, and $\eta_i$, which represents the wage expected agent by agent $i$ from their employer. The parameter $\alpha$ controls the pressure governing wage adjustments.

In Ref.~\cite{WRIGHT2010}, Wright analyzed the emergence of stylized economic facts for $\alpha = 1$. Here, we extend this analysis by examining how the system’s behavior changes as $\alpha$ is varied.
These extensions require a reformulation of the {\it Firing} and {\it Wage payment} rules, which are directly related to wage dynamics. The modified rules are:

\begin{itemize}
\item {\bf Hiring} -- If $i$ is not an employer ($e_i\geq0$):
\begin{enumerate}
    \item A potential employer $j$ is selected from the set $H$  of all agents except employees ($e_j\neq0$), with probability  $P (j)= w_{j}/\sum_{n\in H}w_{n}$.

     \item A wage offer $w$ is drawn uniformly from the interval $\left[\eta_{i},\left(1+\alpha\right)\eta_{i}\right]$, 
     where $\eta_i$ represents the agent's wage expectation and the parameter $\alpha$ therefore controls the intensity of the wage increase sought by the agent. 
     This choice is inspired by Isaac’s analysis of the CSA model, in which he proposes replacing Wright’s original interval $\left[\eta_i, 2\eta_i\right]$ with a fixed incremental adjustment of 1\%, aiming to mitigate the excessively high unemployment rates observed in the original model~\cite{ISAAC2019}. 
     If $w$ exceeds the wealth of the selected employer ($w>w_j$), it is truncated to $w=w_j$.

    \item If the final wage offer exceeds the agent’s expectation ($w > \eta_i$), agent $i$ is hired by employer $j$ ($e_i = j$), and the expectation is updated to $\eta_i = w$. If agent $i$ is not hired and was originally unemployed, a new expectation is drawn uniformly from the interval $\left[0, \eta_i\right]$.

\end{enumerate}

\item {\bf Wage payment and firing} -- If agent $i$ is an employer ($e_i < 0$), we define the set $E$ of employees of $i$ as those agents $j$ for which $e_j = i$. For each employee $j \in E$:

\begin{enumerate}
    \item If employer $i$ has sufficient funds, an amount equal to the employee’s expectation $\eta_j$ is transferred from employer ($w_i \rightarrow  w_i- \eta_j$) to employee ($w_j \rightarrow w_j +\eta_j$), and the employee’s last received wage is updated to $\mu_j = \eta_j$.
    
    \item If employer $i$ does not have sufficient funds, employee $j$ is fired, and their wage expectation is reset to the last received wage, $\eta_j = \mu_j$.  That is, if the agent is fired before receiving the negotiated wage, the expectation is not realized. 

    \item If, after processing the entire set $E$, all employees have been fired, employer $i$ becomes unemployed ($e_i = 0$).

\end{enumerate}

\end{itemize}

\section{Results from agent-based simulations}
\label{sec:results}
 
Observables were collected annually, over 1000 years, after a short transient sufficient to attain a steady state. 
Wealth was calculated as the amount $w_i$ held by agent $i$ at the end of the year, and income $y_i$ as the aggregate wealth received by agent $i$ throughout the year, either as wages for employees, or as revenue for employers. Unless otherwise stated, $N=1000$,  $\overline{w}=2$, $p=1$, which implies that $R=\overline{w}/p=2$, 
and $\Phi_j=\Omega_j=1$, for $j=1,2$. 

\begin{figure}[h!]
    \centering    
    \includegraphics[width = 0.45\textwidth]{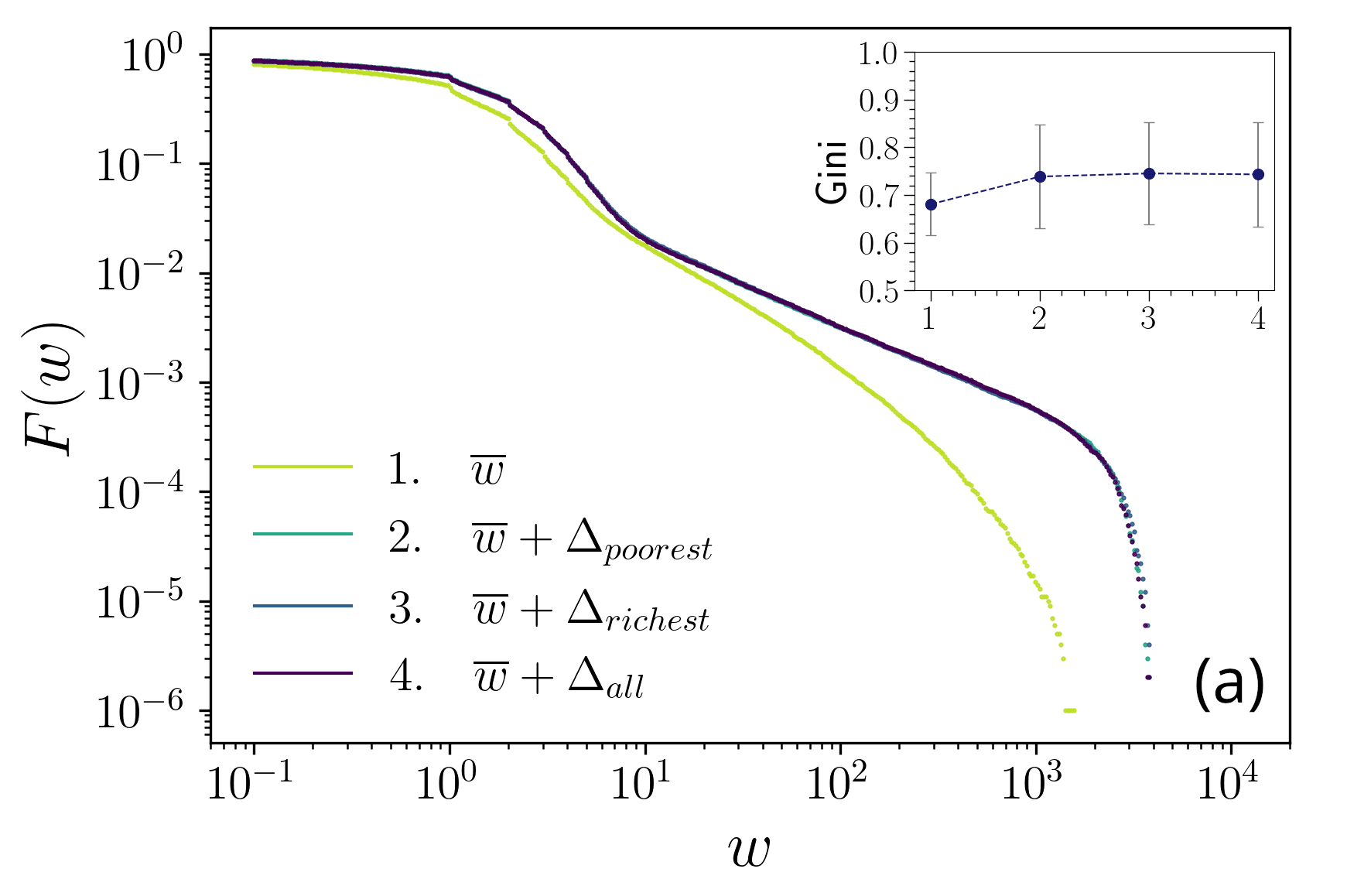}
    \includegraphics[width = 0.45\textwidth]{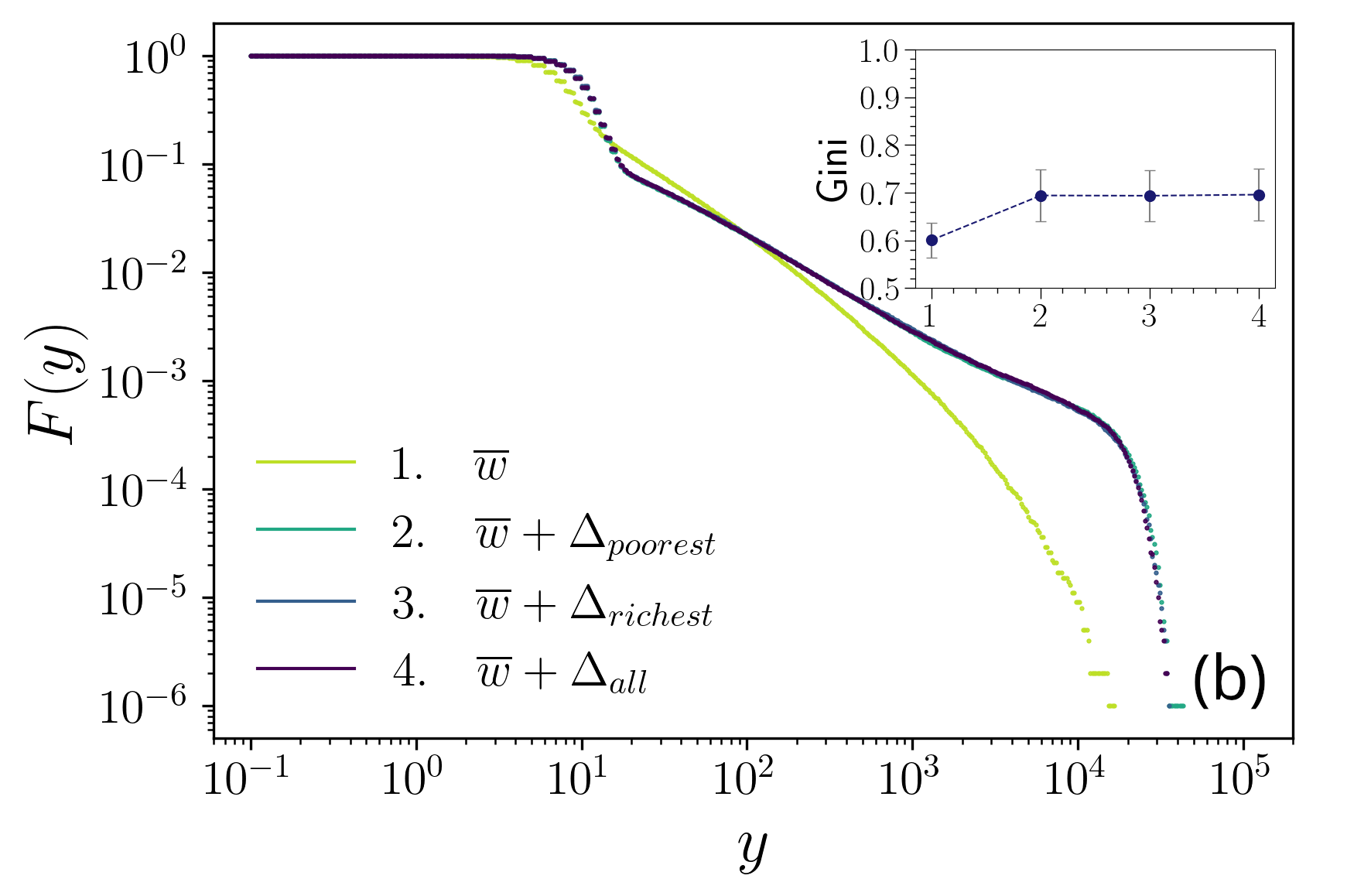}
    \caption{Complementary cumulative distributions of (a) wealth and (b) income  in the steady state for  different preparations of the system: 
    (1) $\overline{w}=2$;
    (2) starting from $\overline{w}=2$, after reaching the steady state, an amount $\Delta=4$ is distributed equally among the poorest 20\% of the agents;
    (3) as in (2), but with the amount distributed among the richest 20\% of the agents;
    (4)   $\overline{w}=2.8$ from the start.  
     The insets show  the Gini index for each scenario.  Note that the distributions are not sensitive to initial condition, but depend only on the wealth per capita $\overline{w}$.
    }
\label{fig:injection}
\end{figure}

We first tested the effect of injecting money into the system, either among the richest or the poorest segments of the population, and found no significant differences compared to the case where the same total wealth is assigned equally from the start. 
The result illustrated in Fig.~\ref{fig:injection} for different protocol of money injection indicates independence on initial conditions. 

\subsection{Expenditure}
\label{sec:expenditure}

\begin{figure}[t!]
    \centering
    \includegraphics[width = 0.45\textwidth]{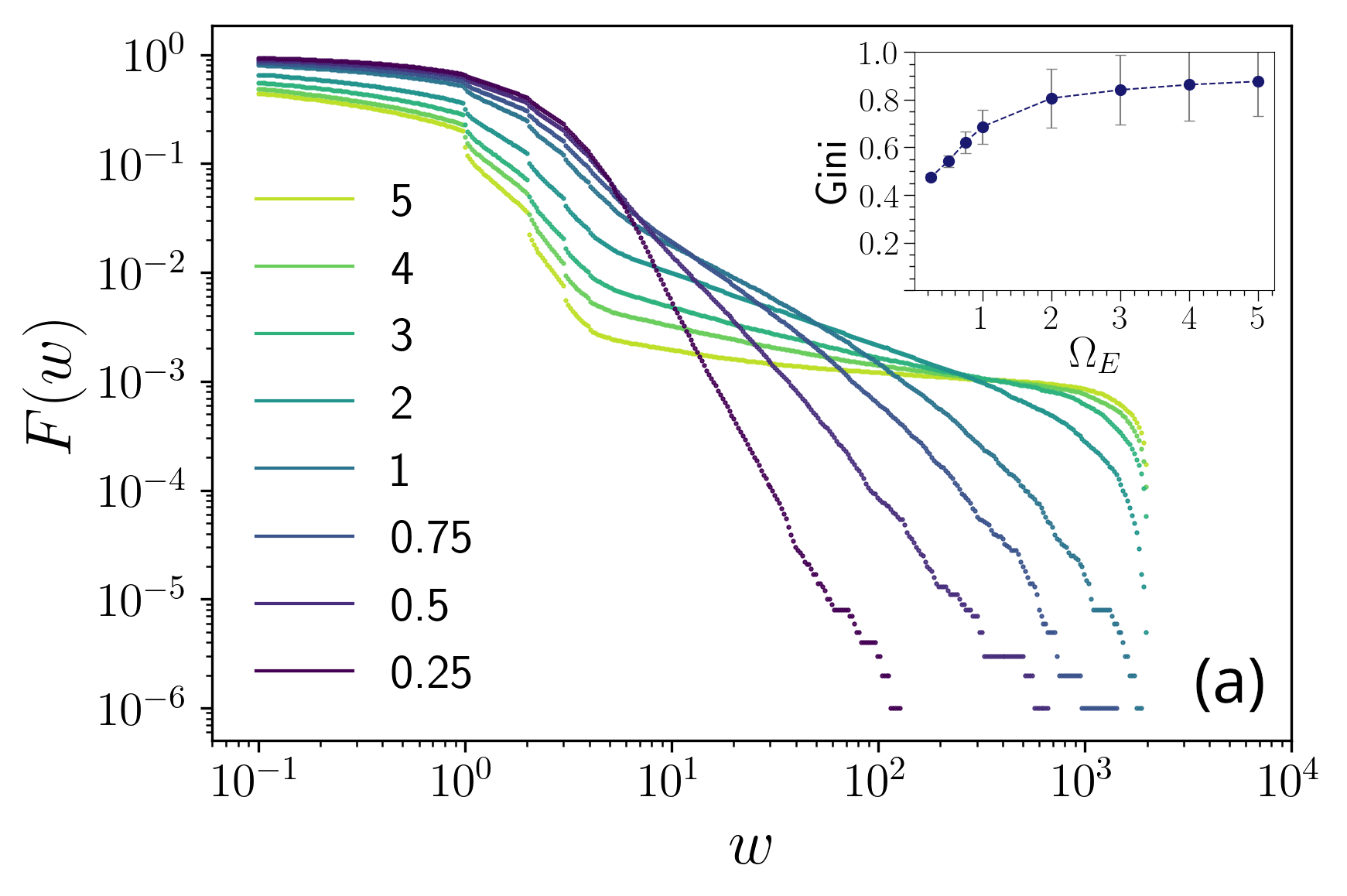}
    \includegraphics[width = 0.45\textwidth]{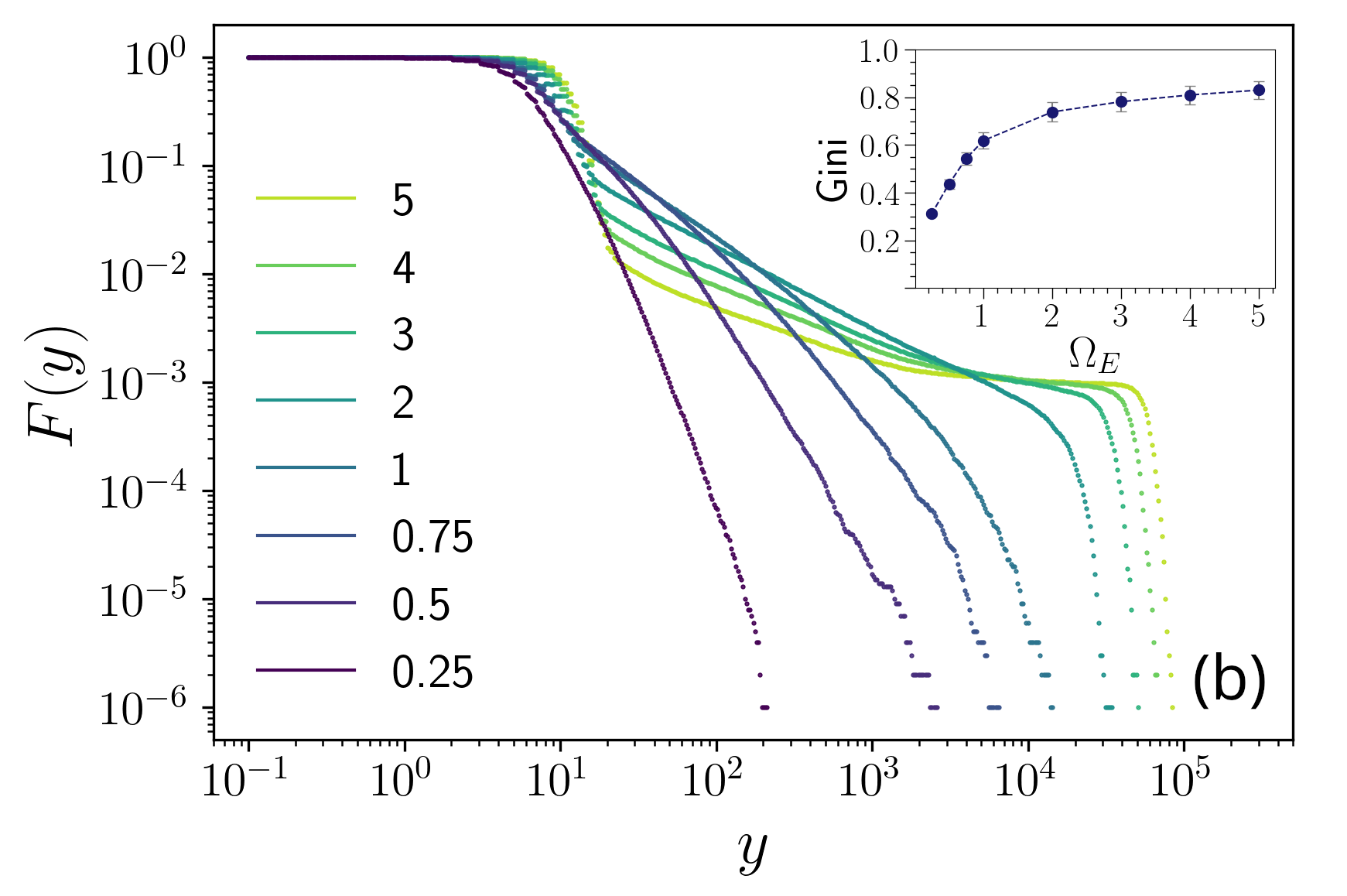}
        \caption{
     Effects of expenditure frequency $\Omega_E$:
Complementary cumulative (a) wealth and (b) income distributions for different values of $\Omega_E$. The insets show the average annual Gini vs. $\Omega_E$. 
 We note that inequality increases as the expenditure frequency increases. 
This is reflected in both the increase in the Gini index and the shift toward heavier tails in the distributions. In the current and subsequent figures, each distribution was built with values recorded at the end of each year, accumulated over $10^3$ years of a single realization. For the inequality indices, we present the average of the annual values (symbols) and their standard deviation (bars).}
     \label{fig:frequency2}
\end{figure}

We examine the effects of varying the frequency $\Omega_E$ with which the {\it Expenditure} rule is applied. The results shown in Fig.~\ref{fig:frequency2} indicate that increasing the frequency with which agents spend in the market leads to higher inequality in both wealth and income. This rise in inequality reflects the emergence and consolidation of monopolistic structures as expenditure becomes more frequent.

Within the model, increasing the frequency of market expenditure results in a larger amount of money being injected into the market value $V$. If the gross domestic product (GDP)---defined here as the aggregate revenue of all firms over a one-year period---is computed as the total firm revenue, then the annual GDP is directly proportional to the total amount of money injected into the market. Consequently, increasing $\Omega_E$ leads to higher GDP.

\begin{figure}[hb!]
    \centering
    \includegraphics[width = 0.45\textwidth]{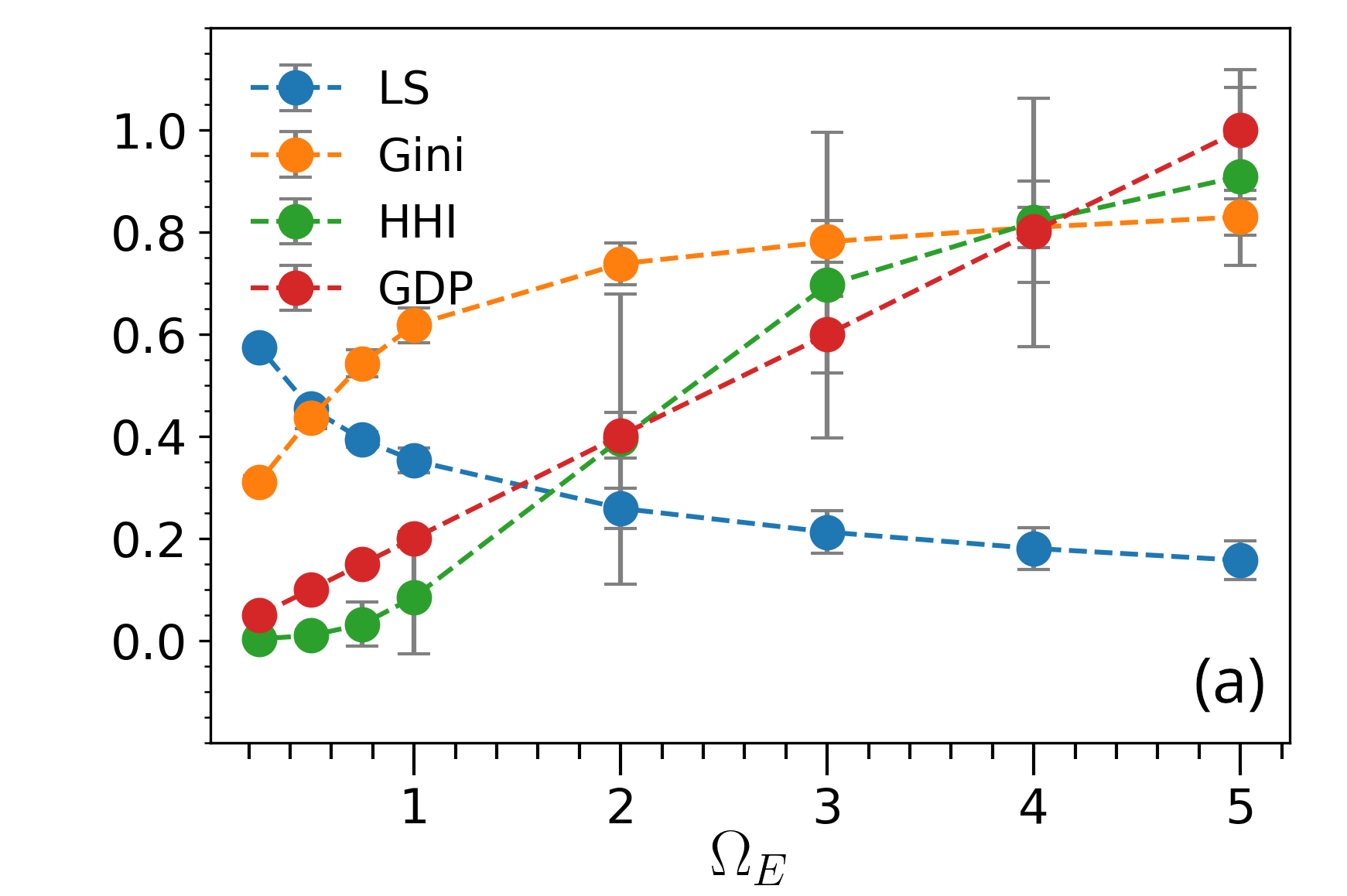}
    \includegraphics[width = 0.45\textwidth]{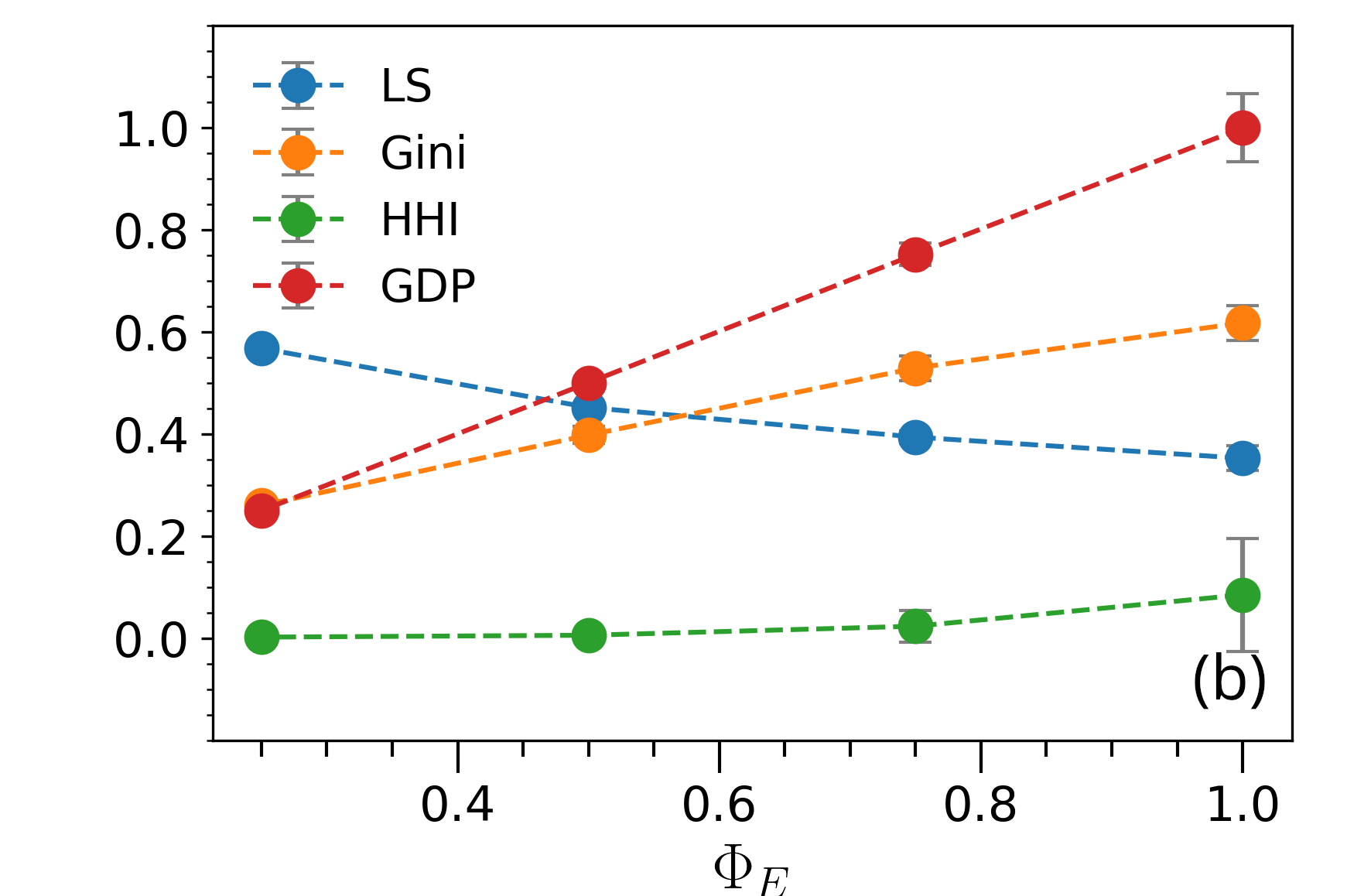}
    \caption{Impact of changes in (a) the expenditure frequency $\Omega_E$ and (b) expenditure fraction $\Phi_E$ over labor share (LS), Gini of income, normalized HHI, and rescaled GDP. 
    The maximum measured value of GDP is $\approx 6\times10^4$ for $\Omega_E=5$ and 
    $\approx 1.2\times10^4$ for $\Phi_E=1$. 
    In contrast with the other quantities, the labor share (LS) decreases with $\Omega_E$ and $\Phi_E$.}
    \label{fig:indices}
\end{figure}

However, while employees’ wages remain fixed at the value $p$, firm income---and therefore employers’ profits---depends on the volume of money circulating in the market. As a result, increasing the flow of money into the market raises GDP but simultaneously alters the composition of income: a larger fraction accrues as firm revenue (profits), while a smaller fraction is paid out as wages. Consequently, the labor share of income---the fraction of total income allocated to workers’ wages---decreases, as illustrated in Fig.~\ref{fig:indices}(a).

\begin{figure}[h!]
    \centering
    \includegraphics[width = 0.45\textwidth]{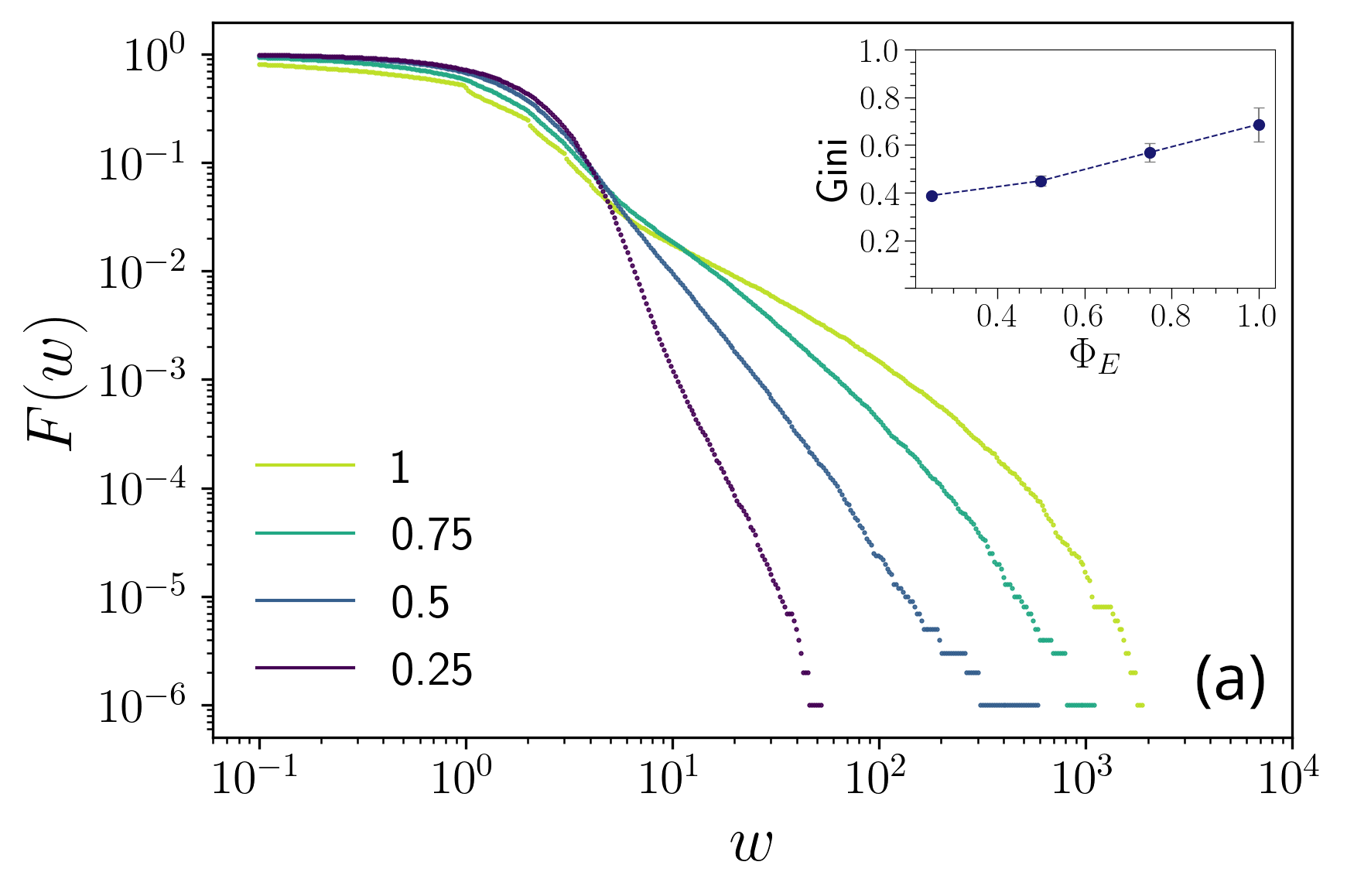}
    \includegraphics[width = 0.45\textwidth]{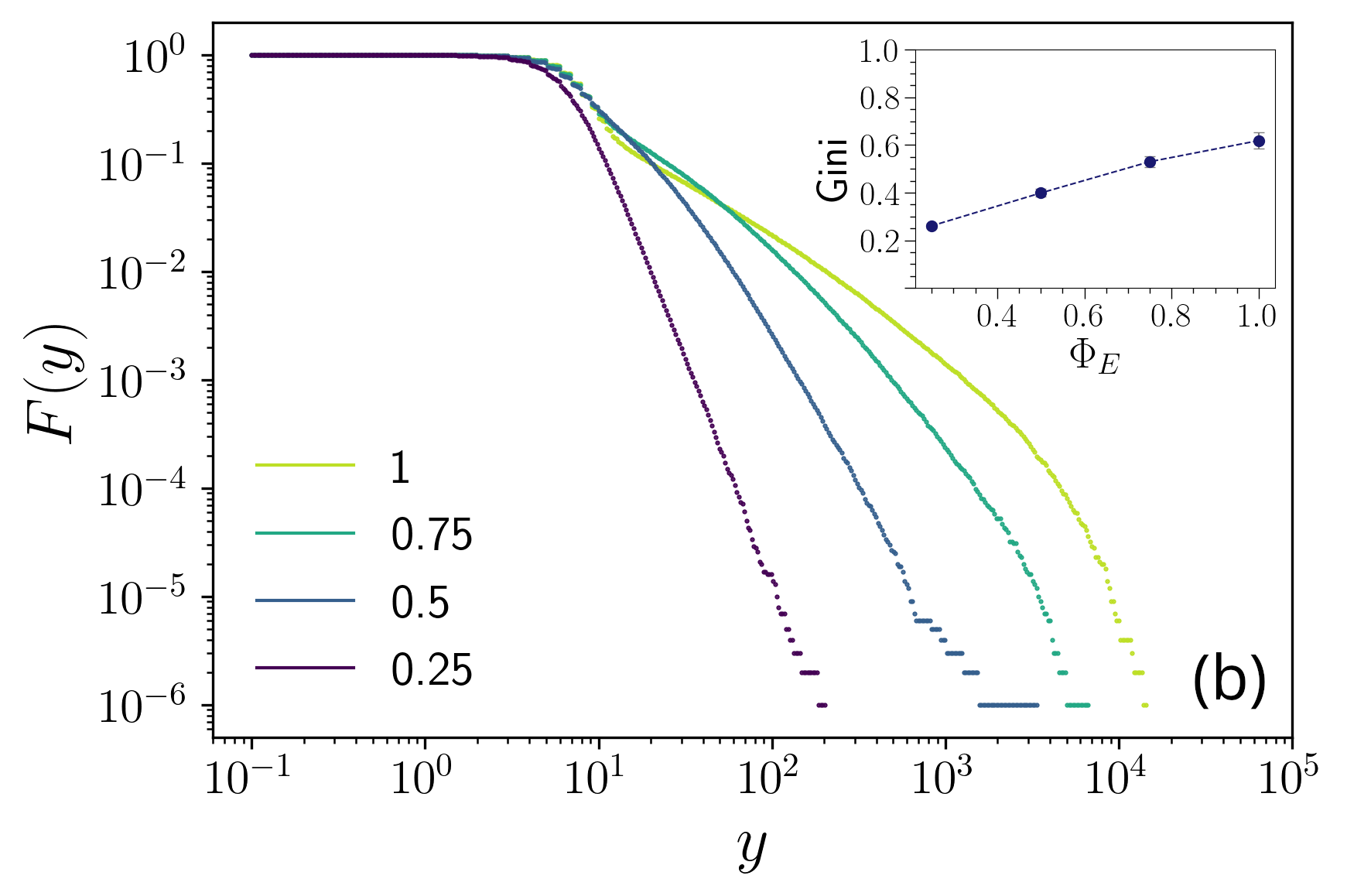}
    \caption{Effects of expenditure fraction  $\Phi_E$:
    Complementary cumulative distributions of (a) wealth and (b) income for different values of $\Phi_E$ indicated in the legend. The inset shows the average annual Gini coefficient vs. $\Phi_E$. The higher the expenditure fraction, the greater the inequality. This is reflected both in the increase of the Gini coefficient and in the shift toward heavier tails in the distributions.}
    \label{fig:fraction2}
\end{figure}

Following similar reasoning, the GDP can also be modified---rather than by changing the frequency $\Omega_E$---by varying the fraction $\Phi_E$ of each agent's wealth withdrawn and deposited into the market value $V,$ as defined in the {\it Expenditure} rule.  The resulting  wealth and income distributions for different values of $\Phi_E$ are shown in Fig.~\ref{fig:fraction2}, with Gini coefficients indicating higher inequality as the fraction $\Phi_E$ increases.

We can gain further insight into the underlying dynamics by examining and comparing different macroeconomic indices represented in  Fig.~\ref{fig:indices}. 
We plot, as a function of $\Omega_E$ (Fig.~\ref{fig:indices}a) and $\Phi_E$ (Fig.~\ref{fig:indices}b), the Gini coefficient of income,  the GDP, the labor share (LS), and additionally the Herfindahl-Hirschman Index (HHI)\cite{resende2009medidas}, a measure of the size of firms in their respective industries and an indicator of competition. 
The index is defined as $H=\sum_{i=1}^L M_i^2$, where $M_i$ is the market share of the firm $i$ and $L$ is the number of firms. Actually, we work with the normalized version of HHI: $H^*=(H-1/L)/(1-1/L)$ if $L>1$, and  $H^*=1$ if $L=1$.  Then $H^*=1$  indicates monopoly while $H^*=0$ indicates perfect competition.

For simplicity, each agent who owned a firm during the year is counted as a single firm, even if it went bankrupt at one point and returned to being capitalist at another within the year. Market share is represented by the fraction of the GDP appropriated by the capitalist. 
All indices are calculated annually at the end of the year and then averaged over 1000 years.

Increasing  $\Omega_E$ and/or $\Phi_E$ raises GDP by allowing more money to circulate in the economy, effectively increasing the total economic output to be distributed among capitalists.  We have also  noted that this increase is accompanied by higher inequality, as measured by the Gini coefficient. 
In addition, there is a tendency toward the formation of monopolies, a behavior captured by the increasing  HHI. But perhaps the most visible effect is on the labor share. Changing the {\it Expenditure} rule in order to ‘heat up’ the economy and increase GDP, under these conditions, leads directly to a reduction in the labor share, decreasing the the fraction of total income accruing to workers and thereby further increasing  inequality.

\subsection{Market revenue} 
\label{sec:revenue}

\begin{figure}[h!]
    \centering
    \includegraphics[width = 0.45\textwidth]{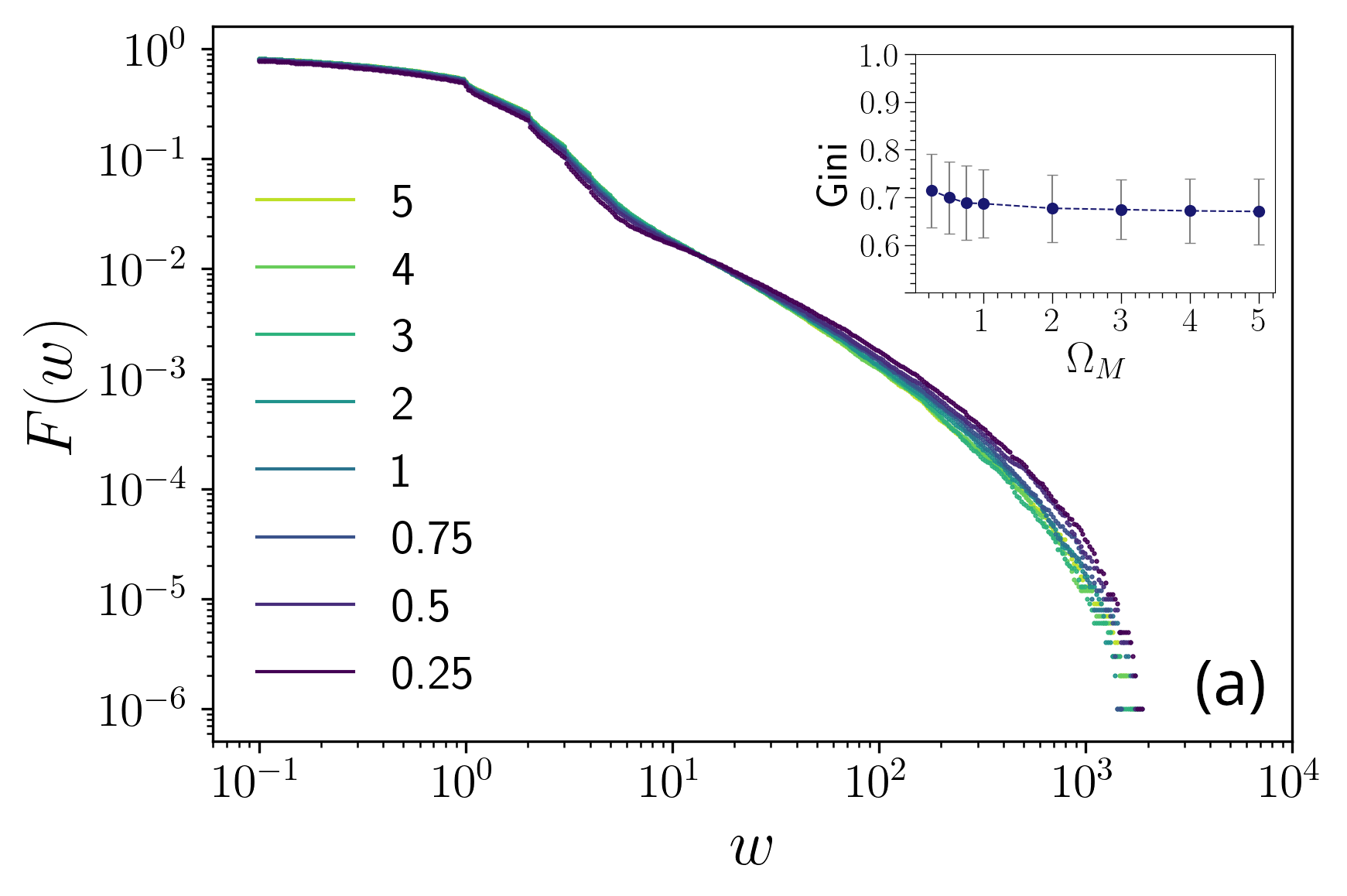}
    \includegraphics[width = 0.45\textwidth]{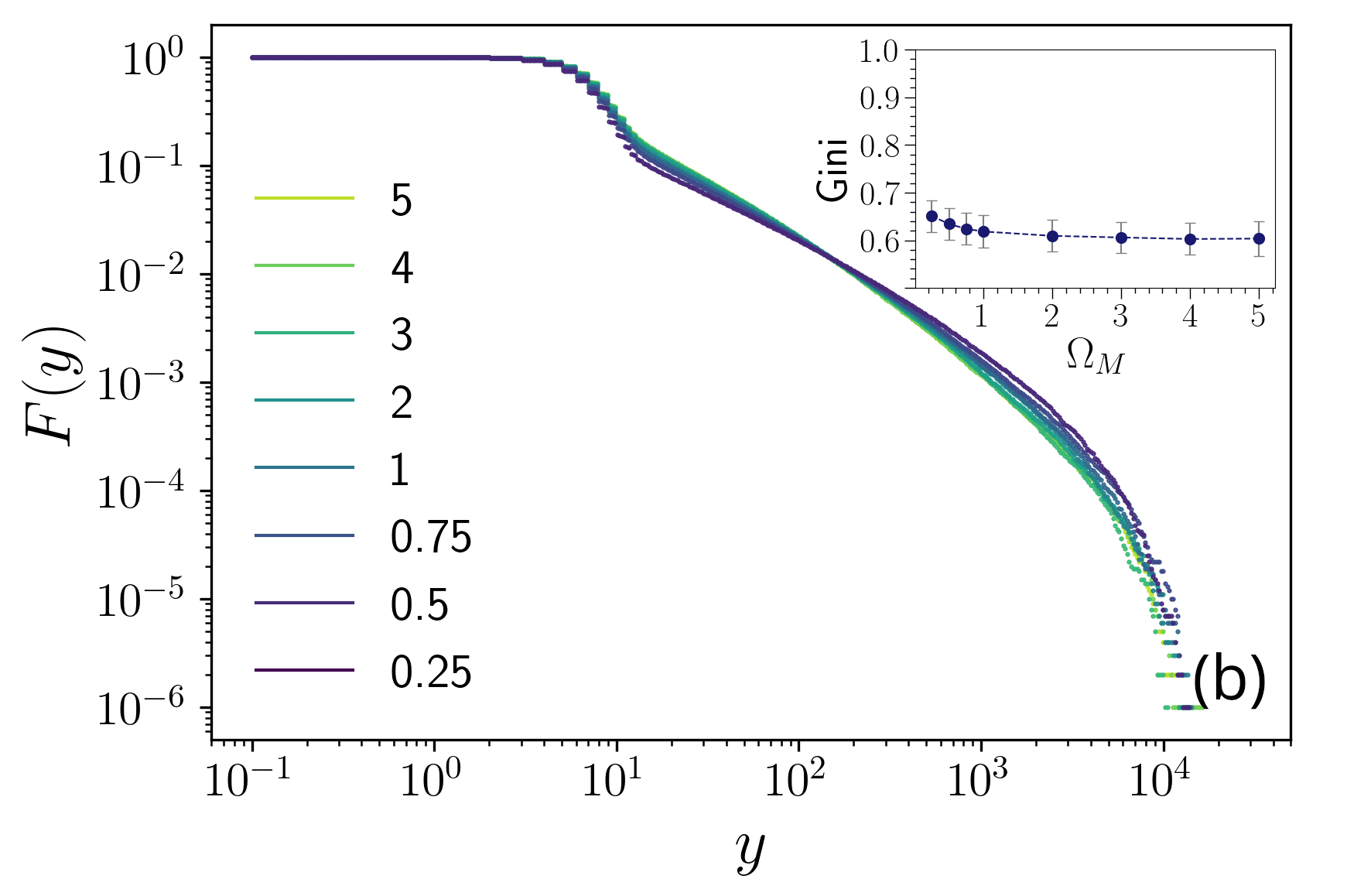}
        \caption{
        Effects of market revenue frequency $\Omega_M$: complementary cumulative (a) wealth and (b) income distributions for different values of $\Omega_M$. The inset shows the average annual Gini  coefficient as a function of $\Omega_M$, indicating that inequality decreases slightly as the frequency of revenue withdrawals increases, with saturation for $\Omega_M > 1$.
        }
     \label{fig:frequency3}
\end{figure}

We now analyze the modifications of the {\it Market Revenue} rule. As we can see in Fig.~\ref{fig:frequency3}, the effect of changing the frequency $\Omega_M$ is more subtle and opposite to that observed for $\Omega_E$. That is, increasing the $\Omega_M$ softly decreases the inequality measured by the Gini coefficient.

\begin{figure}[h!]
    \centering
    \includegraphics[width = 0.45\textwidth]{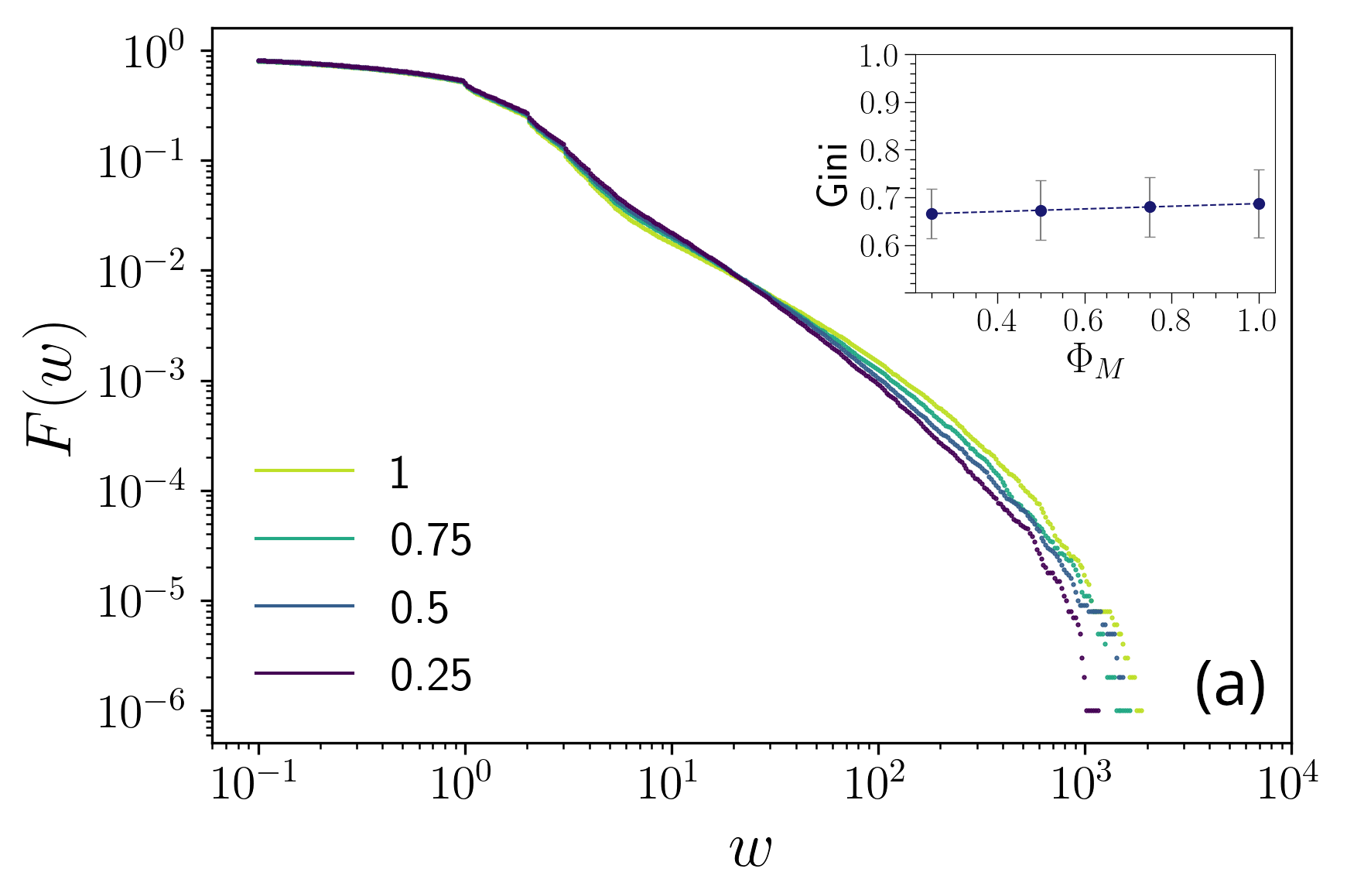}
    \includegraphics[width = 0.45\textwidth]{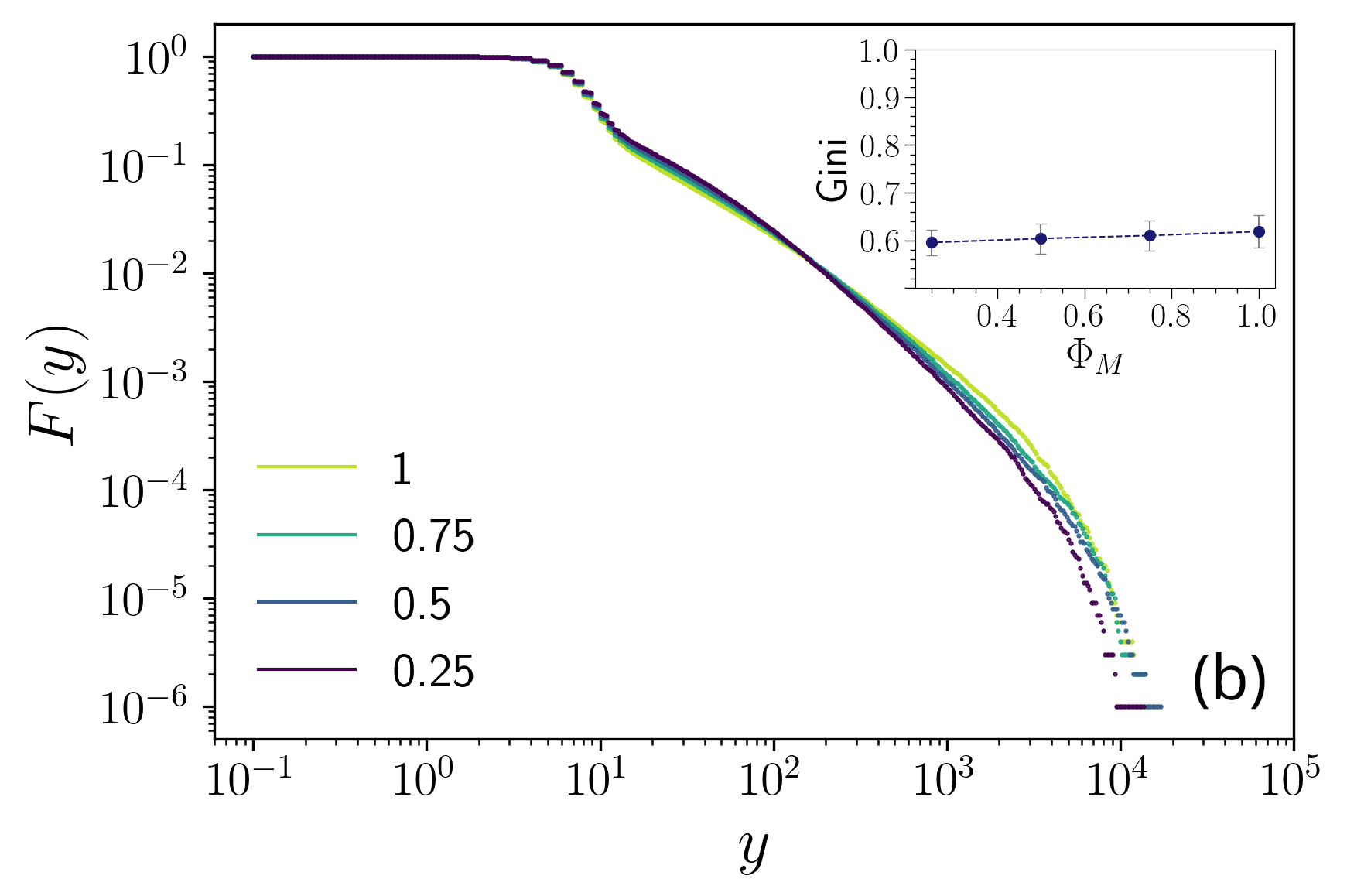}
    \caption{Effects of market revenue fraction $\Phi_M$: complementary cumulative (a) wealth and (b) income distributions for different values of $\Phi_M$. The inset shows the average annual Gini 
     coefficient as a function of $\Phi_M$, indicating a nearly negligible increase in inequality as the fraction of wealth withdrawn from the market increases.}
    
    \label{fig:fraction3}
\end{figure}

When we analyze the effects of changing $\Phi_M$, that is, the fraction withdrawn from the market in the {\it Market revenue} rule,   we observe in Fig.~\ref{fig:fraction3},  a similar effect to that observed when varying the value of $\Phi_E$, the inequality increases again, but more subtly. Interestingly, increasing $\Phi_M$ and $\Omega_M$ produces opposite effects on the system's inequality.

Unlike the changes in the {\it Expenditure} rule,  now the GDP remains unchanged (not shown). That is, modifications in how money is withdrawn from the market value do not affect consumption, therefore keeping GDP constant. Since, in this model, GDP is  given by the aggregate annual income of capitalists, this also implies that the total income appropriated by capitalists does not change. The labor share likewise remains constant, meaning that the fraction of total annual income allocated to workers’ wages is unchanged.

Although these indices remain constant, a subtle but systematic variation in the Gini index can be observed across all figures corresponding to modifications of the {\it Market Revenue} rule. Specifically, inequality increases with $\Phi_M$ and decreases with $\Omega_M$. 
This pattern indicates that the observed change in inequality must originate from variations within classes rather than between them. Since wages are fixed in the model and the total income allocated to workers remains unchanged, the variation in inequality must therefore arise within the capitalist class.

This effect is reflected in the behavior of the HHI, which closely tracks the variation of the Gini coefficient. Increasing $\Phi_M$ raises the fraction of market value appropriated by firms at each step; as a result, there is a slight increase in monopoly power and, consequently, in the Gini coefficient.  
This mechanism becomes particularly pronounced when there has been a large deposit that makes the market value $V$ exceed its average level. 
In this case,  increasing $\Phi_M$ implies that a larger fraction of the accumulated value is withdrawn each time the {\it Market revenue} rule is applied, and less of this value will be distributed among capitalists, increasing the concentration of GDP distribution and, consequently, the concentration of wealth. In extreme cases with $\Phi_M=1$, the entire market value could be appropriated by a single capitalist.

It is worth noting that, regardless of the value of $\Phi_M$, GDP does not increase. This means that it does not change consumption in the system or the total money available in market value to be appropriated by capitalists throughout the year. 

Now, when we increase $\Omega_M$, we observe a more pronounced effect on the variation of inequality for $\Omega_M\leq 1$. In this case, the lower the frequency, the more often agents can deposit money in the market through consumption before the agent withdraws it, increasing the total amount of money available for withdraw. 
So now, the effect of varying the $\Omega_M$ frequency no longer depends as much on above-average deposits, its influence is more strongly felt in the aggregate of ordinary transactions, especially depending on the amount of market value withdrawals that exist between one deposit and another.

Remember that we now maintain $\Phi_M=1$, which means the worker can withdraw all the money available in the market at once. Thus, we can hypothesize that increasing the frequency has the effect of preventing the accumulation of money in the market and better redistributing revenue among firms, slightly reducing the level of monopoly and, consequently, the Gini index. For $\Omega_M>1$, workers go to the market more than once before new consumption occurs, so there is always less money available with each subsequent visit.  

We can consider an extreme situation where a worker withdraws all the money at market value the first time, regardless of how many other withdrawal attempts are executed before new consumption occurs, no revenue will be possible, and we would have the same practical effect as if we had $\Omega_M=1$. Therefore, increasing the frequency of application of rule 3 for values above 1 tends to have a less significant effect than that for values below 1.
 
More precisely, both changes alter the distribution of GDP among capitalists without altering the total quantity. In the first case,  when we change the fraction, we have the same amount of money distributed among capitalists with the same amount of withdrawals from the market. But for lower $\Omega_M$ values, the distribution of GDP among capitalists is more uniform; increasing the fraction increases the number of higher and lower value withdrawals from the market value, making it more unequal and slightly affecting the overall inequality of the system. 

In the second case, by changing the frequency of withdrawals from the market, we alter the average amount of money withdrawn, with more significant effects for $\Phi_M \le 1$, as it allows the accumulation of deposits at market value between each withdrawal attempt , since for $\Phi_M>1$ there is more than one withdrawal attempt before a new deposit is made at market value. Since total GDP remains constant, fewer visits to the market imply larger withdrawals and a more concentrated profit distribution.

\subsection{Hiring, wage payment and firing rules} 
\label{sec:hiring}

Wright reports that the CSA model presents an unemployment rate of 18.5\%, a value considered higher than what is normally reported by modern economies. This value is originally compared to expanded measures of unemployment, as more traditional measures only take into account people who are unemployed and actively looking for work. A.G. Isaac provides the example that during the Great Recession in the US, the U6 unemployment rate reached just over 17\%~\cite{ISAAC2019}.  
This value accounts for both active job seekers and underemployed or discouraged workers. 

A.G. Isaac proposes some modifications aimed at correcting minor implementation errors and making the results more realistic. Among the changes introduced, the one of particular interest here is the reduction in the pressure for wage increases, controlled by $\alpha$, as defined in Section \ref{modifications}, from $\alpha=1$ to $\alpha=0.01$. In both studies, the authors examined only a single fixed value of $\alpha$.

Lin Lin also draws attention to the problem of high unemployment in Wright's original model and proposes a modified version that achieves a low unemployment rate~\cite{lin-lux}. Due to differences in the rules, there is not a variable parameter similar to  $\alpha$, but only one scenario is analyzed,  namely, a single level of pressure for wage adjustments by workers.

\begin{figure}[h!]
    \centering
    \includegraphics[width = 0.48\textwidth]{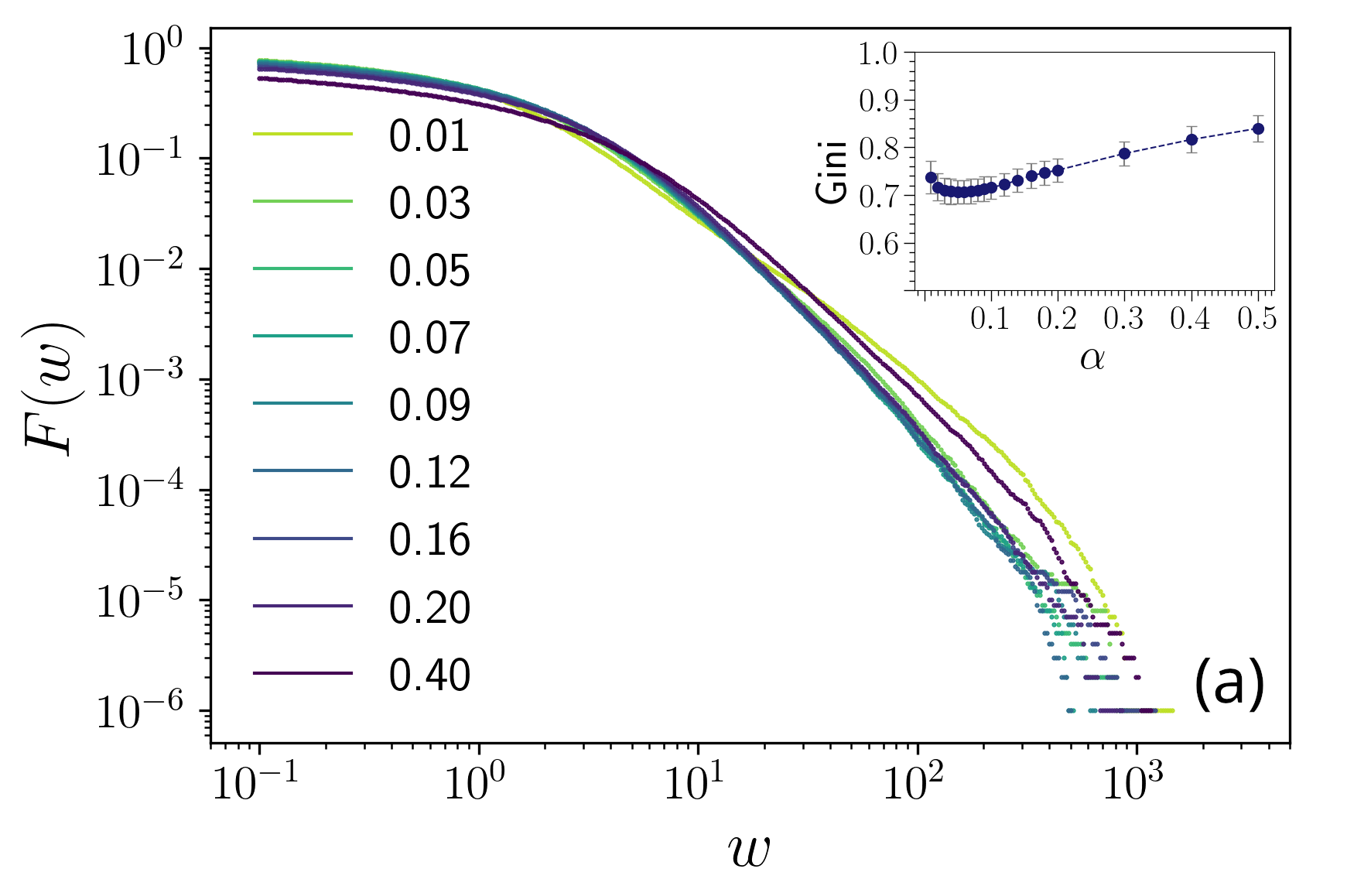}
    \includegraphics[width = 0.48\textwidth]{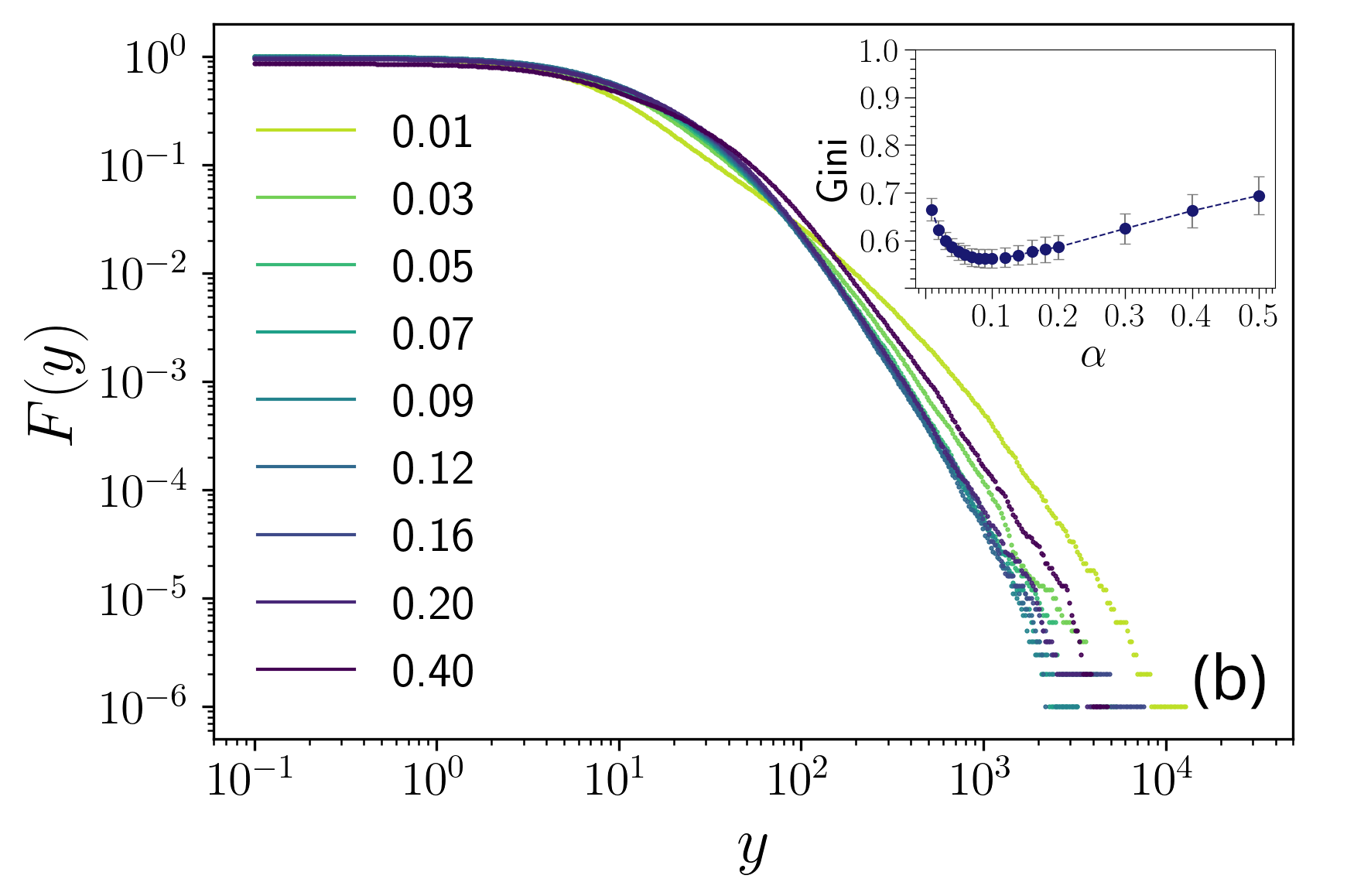}    
    \caption{Complementary cumulative (a) wealth and (b) income distributions for different values of $\alpha$  (wage bargaining intensity parameter).    The insets show  the average annual Gini coefficient as a function of $\alpha$.  The dependence of inequality on workers’ wage bargaining power is non-monotonic: inequality initially decreases, reaches a minimum, and then increases, eventually returning to values comparable to or exceeding those at $\alpha \approx 0$.}

 \label{fig:reajuste}
\end{figure}

First, we verified that increasing the per capita wealth  $\overline{w}$ leads only to a rescaling of the distributions, leaving $R$ unchanged. Therefore, to explore variations in $R$, we directly modify the wage-bargaining parameter $\alpha$.  Figure~\ref{fig:reajuste} shows how the wealth and income distributions evolve as $\alpha$ varies. 
Inequality displays a nonmonotonic dependence on $\alpha$, with a minimum around $\alpha\simeq5\%$ for wealth and $\alpha\simeq 10\%$ for income, even though the average wage rises steadily with $\alpha$.  
By contrast, the labor share increases throughout the explored range of $\alpha$, 
while GDP and HHI remain largely insensitive to this parameter(not shown).

\begin{figure}
    \centering
    \includegraphics[width = 0.48\textwidth]{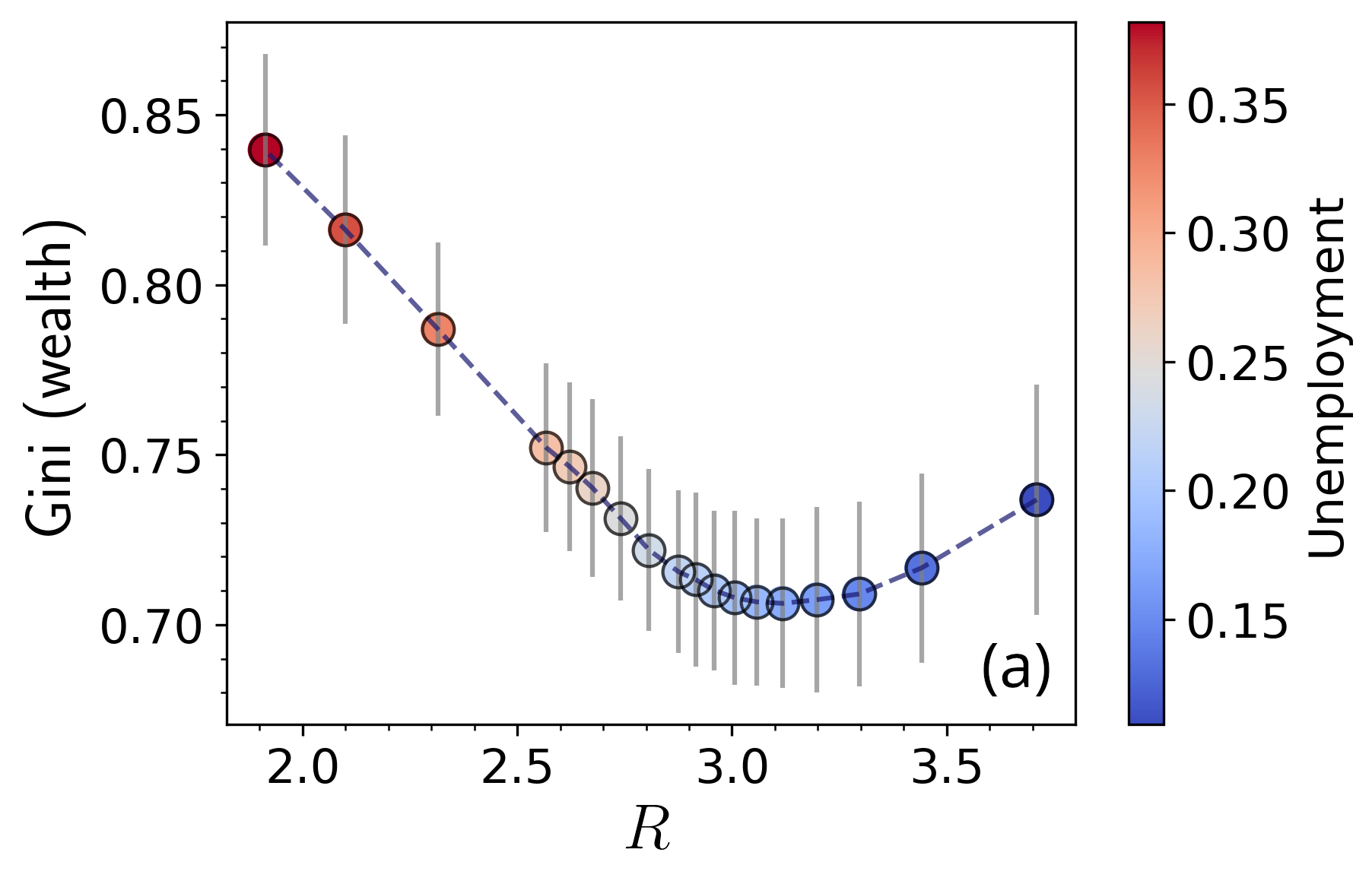}
    \includegraphics[width = 0.48\textwidth]{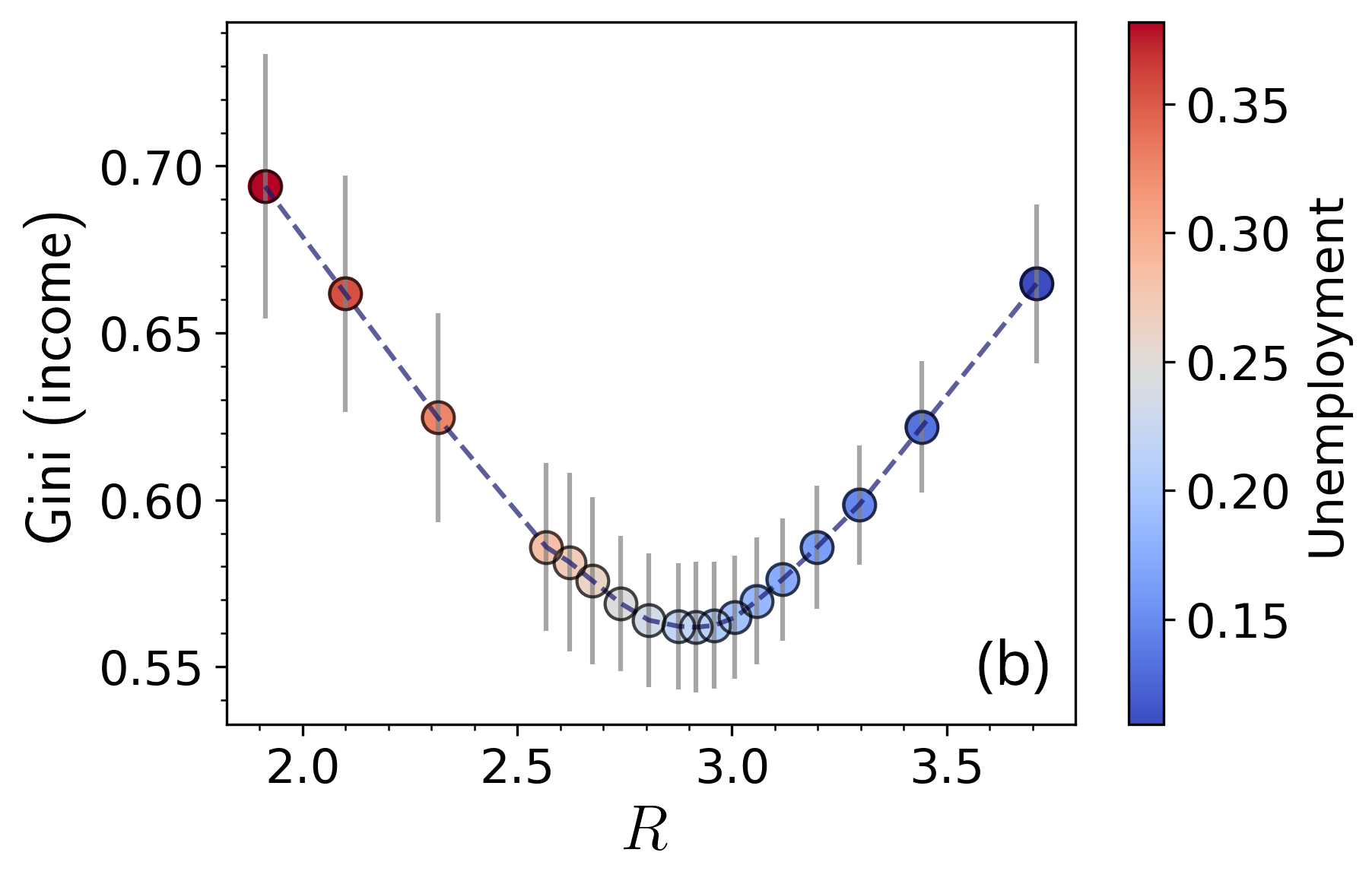}
    \caption{
    Gini coefficient as a function of $R$ (wealth-to-wage ratio). 
    The different points correspond to different values of the average monthly wage, which in turn are obtained by varying the bargaining intensity parameter, $\alpha$. The color of the point indicates the unemployment rate that emerges from each $\alpha$. Uncovering the unemployment rate helps clarify the non-monotonic behavior of the Gini coefficient: for unrealistically high unemployment, inequality decreases as $R$ increases; as the unemployment rate approaches more realistic levels, inequality increases with further increases in $R$.
}
 \label{fig:desemprego2}
\end{figure}

To understand the nonmonotonic behavior of the Gini indices, we examine Fig.~\ref{fig:desemprego2}, focusing on the unemployment rate. 
First, recall that increasing $\alpha$ raises the average wage $\overline{p}$, which in turn reduces the ratio $R=\overline{w}/\overline{p}$. Thus, $R$ and $\alpha$ are inversely related. 
For large enough values of $R$ (approximately $R\gtrsim 3$), the model produces realistic unemployment rates (approximately $18\%$, from the color scale in the figure), displaying increasing inequality as $R$ increases. 
This behavior is consistent with results originally obtained under constant wages and is also observed in empirical data.

This behavior can be explained as follows. As $R$ increases, the fraction of total wealth disputed by capitalists over the year—measured by the GDP—remains constant. However, we observe that both the labor share and the average annual wage per agent decrease, 
enabling capitalists to hire more agents, thereby lowering unemployment. Nevertheless, the resulting increase in employment, together with the reduction in the size of the capitalist class, is insufficient to compensate for falling wages, ultimately leading to higher inequality and greater monopoly concentration.

 The most challenging regime to interpret corresponds to low values of $R\lesssim 3$, where we obtain the novel result that increasing $R$ leads to a reduction in inequality. This behavior can be understood by examining class sizes. At low  $R$,
the initially high unemployment rates imply that, at any given time, an unrealistically large fraction of the population has no income. In this regime, although increasing $R$ lowers the average wage of employed agents, it simultaneously enhances the capacity of capitalists to hire additional workers, leading to a decline in unemployment toward more realistic levels.

This reduction in unemployment, together with the expansion of the capitalist class, promotes a substantial redistribution of wealth across the system and ultimately results in lower inequality. Notably, the size of the capitalist class exhibits a non-monotonic dependence on $R$, varying inversely with the behavior of the Gini coefficient.
 
These results are also relevant for assessing the validity of the model when compared with empirical unemployment data.

\section{Conclusions and final remarks}
\label{sec:final}

It was already known that in this kind of model, with fixed wages, increasing per capita wealth (and consequently total wealth) leads to a concentration of wealth and income, which indicates that the wealth added to the system is not distributed equally but is concentrated in the capitalist class. By the way, this wealth concentration is a common feature of wealth exchange models, such as the Yard-Sale  model and its variants~\cite{yakovenko2023,chakraborti2011,boghosian2014a,chakrabarti2013,boghosian2014}.  By combining the most recent results, we can take a step further and understand the effects of other interferences.

Increasing each agent's spending injects more money into the market value. The component of income that depends directly on the size of market value is capitalists’ profit revenue. 
Since employees have a fixed wage, modifications  to {\it Expenditure } rule that increase GDP imply an increase in capitalists' income and, consequently, a worsening of inequality. 
In other words, as with total wealth, an increase in GDP does not necessarily lead to a more equitable distribution of wealth within the system; on the contrary, it results in higher inequality. 
In this model, such growth leads to a decline in the labor share, meaning that an increasingly larger fraction of total income is appropriated by capitalists. It also intensifies monopoly power, increasing inequality even within the capitalist class, an effect captured by the HHI.
This result is consistent with the central argument advanced by Piketty~\cite{piketty}, who, based on extensive historical and empirical evidence, shows that economic growth alone does not necessarily translate into a more equitable distribution of wealth,particularly when growth is not accompanied by proportional increases in labor income. 

Modifications to {\it Market revenue} rule do not change GDP or market value, since wages are fixed. Since market value is the money that capitalists compete for at a given moment and GDP is defined as the money that is withdrawn from market value during a year,  GDP is thus the money that capitalists compete for throughout the year. Thus, while changes in {\it Expenditure} rule increased GDP, leading to an increase in the money appropriated by capitalists, changes in {\it Market revenue} rule only change how it is distributed, altering the inequality within the capitalist class. Thus, the effects of these changes on the inequality of the entire system are more subtle.  

Finally, if we release the system from the fixed wage constraint, total wealth becomes irrelevant because wealth and income distributions scale with total wealth. However, changing the bargaining intensity parameter ($\alpha$) leads to a change in the average wage paid to workers. As the average wage increases, unemployment also increases, a feature that becomes extremely important here. For the parameter range in which the system has an unemployment rate close to real values ($1\%\leq\alpha\leq5\%$), we reproduce the same effect observed for fixed wages. That is, as we increase the average wage, with fixed per capita wealth (i.e., we decrease $R=\overline{w}/p$), inequality declines. However, above a certain threshold, the unemployed class becomes excessively (and unrealistically) large, and inequality eventually rises again.

{\bf Acknowledgments:}
We are grateful to Victor Yakovenko for stimulating discussions at the 2025 Econophysics Colloquium. 
We acknowledge financial support from CAPES (Fundação Coordenação de Aperfeiçoamento de Pessoal de Nível Superior), through finance code 001, 
and CNPq (Conselho Nacional de Desenvolvimento Científico e Tecnológico) for Universal 406820/2025-2. 
C.A. also acknowledges partial financial support from CNPq for grant 308347/2025-0) and FAPERJ (Fundação de Amparo à Pesquisa do Estado do Rio de Janeiro), for CNE E-26/204.130/2024. S.G. would also like to thank CNPq for partially supporting this work under grant 309560/2025-0.

{\bf Data availability:}
%Data were obtained from numerical simulations and are available under request.
Data were obtained solely from numerical simulations performed with the programs available at \cite{code}.

{\bf Author Contributions:}
JSB: investigation, methodology, formal analysis, writing.
CA:  supervision, conceptualization,  investigation, methodology, writing, funding acquisition.
SG:  supervision, conceptualization,  investigation, methodology, writing, funding acquisition.

\bibliographystyle{unsrt}
\bibliography{referencias}
\end{document}